\documentclass[11pt]{article}
\usepackage{fancyhdr}
\usepackage{url}
\newcommand{\cancel}[1]{}

\usepackage{amsfonts}
\usepackage{color, float}
\usepackage{appendix}
\usepackage{amsmath,amsthm}
\usepackage{amssymb, longtable}
\usepackage{graphicx,xypic,tikz}
\usepackage{algorithm}
\usepackage{algorithmicx}
\usepackage{algpseudocode}
\usepackage{url,hyperref,enumerate}
\usepackage{float}

\usepackage{authblk}
\usepackage{hyperref}

\usepackage{amsmath,stackrel}

\usepackage{url}

\usepackage{cite}
\usepackage{cleveref}
\usepackage{array}
\usepackage{tikz}
\usepackage{wrapfig}
\usetikzlibrary{decorations.markings}
\usetikzlibrary{shapes.geometric}
\usetikzlibrary{positioning}

\usetikzlibrary{crypto.symbols}

\usetikzlibrary{shadows}
\usetikzlibrary{decorations.pathmorphing}
\usepackage{amsfonts,euscript,fancybox,color}

\usetikzlibrary{shapes,arrows,automata}
\usetikzlibrary{calc}
\usetikzlibrary{shapes,arrows}

\usepackage{graphicx, psfrag,fancyhdr,layout,subfigure}
\usepackage{listings}
\usepackage{amsmath,amssymb, enumerate}

\usepackage{epsfig}
\usepackage{amsfonts}
\usepackage{graphicx}
\usepackage{color}

\usepackage{bbm}
\usepackage{amscd}

\usepackage{multirow}
\usepackage{multicol}

\newsavebox{\savepar}

\newcommand{\F}{{\mathbb F}}

\newcommand{\sliscp}{{\sf sLiSCP}{}}

\newcommand{\mb}{\mathbf}
\newcommand{\Fn}{\mathcal{F}}
\newcommand{\kdf}{\texttt{KDF}}
\newcommand{\mic}{\texttt{MIC}}

\newcommand{\ace}{\textsf{ACE}}
\newcommand{\spix}{\textsf{SPIX}}
\newcommand{\wage}{\textsf{WAGE}}

\newtheorem{construction}{Construction} 

\usepackage{threeparttable}

\usepackage{geometry}
\title{\bf Implementation of Three LWC Schemes in the WiFi 4-Way Handshake with Software Defined Radio}
\author[1]{Yunjie Yi}
\author[1]{Guang Gong}
\author[2]{Kalikinkar Mandal\thanks{The work was done when Kalikinkar Mandal was with the Department of Electrical and Computer Engineering, University of Waterloo, Canada.
}}
\affil[1]{Department of Electrical and Computer Engineering, University of Waterloo, Waterloo, ON, N2L 3G1, Canada\\
\texttt{\{y22yi,ggong\}@uwaterloo.ca}
}
\affil[2]{Faculty of Computer Science, University of New Brunswick, Fredericton, NB, E3B 5A3, Canada\\ \texttt{kmandal@unb.ca}
}
\date{}

\renewenvironment{abstract}
{\small\quotation {\bfseries\noindent{\large\abstractname}.} }

\begin{document}

\pagenumbering{arabic}

\maketitle

\begin{abstract}
With the rapid deployment of Internet of Things (IoT) devices in applications such as smarthomes, healthcare and industrial automation, security and privacy has become a major concern. 
Recently, National Institute of Standards and Technology (NIST) has initiated a lightweight cryptography (LWC) competition to standardize new cryptographic algorithm(s) for providing security in resource-constrained environments. In this context, measuring the suitability of new algorithms with existing communication and authentication protocols is an important problem. This paper investigates the performance of three NIST lightweight authenticated ciphers in round 2 namely {\ace}, {\spix} and {\wage} in the WiFi and CoAP handshaking authentication protocols. We implement the WiFi and CoAP handshake protocols and the IEEE802.11a physical layer communication protocol in software defined radio (SDR) and embed these two handshaking protocols into the IEEE802.11a OFDM communication protocol to measure the performance of three ciphers. We present the construction of {\kdf} and {\mic} used in the handshaking authentication protocols and provide optimized implementations of {\ace}, {\spix} and {\wage} including  {\kdf} and {\mic} on three different (low-power) microcontrollers. The performance results of these three ciphers when adopted in WiFi and CoAP protocols are presented. Our experimental results show that the cryptographic functionalities are the bottleneck in the handshaking and data protection protocols. 
\end{abstract}

\noindent\textbf{Keywords.} Internet of Things (IoT), Security and privacy, Lightweight cryptography, Microcontroller implementation, Authentication protocol, IEEE 802.11a OFDM transmission, Software defined radio


\section{Introduction}

With the rapid growth of the Internet of Things (IoT), it penetrates into our daily life deeply and poses extraordinary effects on us. The IoT connects a wide range of devices, spanning from tiny smart devices to computers and servers. Most of those IoT devices such as sensors, actuators and radio frequency identification (RFID) tags are wirelessly connected though Internet, bluetooth, vehicular ad-hoc networks (VANETs) and equipped with microcontrollers  and radio frequency (RF) transceivers. They communicate with each others to collect various types of data from applications such as industrial and building control, e-health (e.g., medical devices embedded in our body or skin),   smart home (e.g., lights, TV, thermostats, cameras, washing machines, dryers, and refrigerators), smart grid, self-driving cars, and other embedded systems. 
%
According to experts, the growth rate of IoT is approximate 20\% per year, and  the greatest challenges for the IoT ecosystems are security, scalability, and reliability~\cite{iot-predictions}.
In 2017, the number of IoT devices has crossed the world's population, and it is expected to be around 75 billion connected IoT devices by 2025 \cite{forbs}. 
In order to reach its portability and small size, most of IoT devices have limited computational ability and limited power supply.
According to current developments, the most rapid growing applications of IoT are smart cities (approx. $26\%$), industrial IoT (approx. $ 24\%$), connected health (approx. $20\%$), smarthomes (approx. $14\%$), connected cars, wearable devices, and smart utilities \cite{gong18}.

As an IoT system consists of heterogeneous devices, the devices are connected through different types of wireless communication protocols. The major organizations such as IEEE and IRTF for standardizing  communication and security protocols  have moved to support IoT systems. In newly amended IEEE 802.11ax for WiFi systems \cite{IEEESA}, it targets at supporting  established frequency bands with low power and low complexity operations, meaning it may support  the access point (AP) to interact to the client device (and vice versa) at data rates as low as 375 Kbps \cite{ieee802.11ax}. 
On the other hand, new lightweight protocol standards such as MQTT and CoAP \cite{rfc7252,mqtt}  and  MAC protocols \cite{Bird-MAC,MAC-IoT} for tiny IoT devices are developed while considering the factors such as limited resources, communication patterns, and interoperability. 
Recently several protocols and key generation techniques have been developed \cite{iTLS,LightAKA,MobiKey}. 
In \cite{iTLS}, a lightweight secure transport protocol is proposed which provides implicit mutual authentication. In \cite{LightAKA}, a lightweight authentication with key agreement protocol is developed for smart wearable devices. A key generation technique for symmetric-key ciphers is proposed based on the channel feature for smart home applications \cite{MobiKey}. 
The upcoming cellular 5G system has aimed at enabling IoT for connecting a growing number of cars, meters, machinery sensors, etc \cite{5GPPP}.  

While the value of user data has given rise to new economic opportunities such as data markets, it generates new vulnerabilities for security and privacy caused by cyber attacks. 
In recent years, several attacks have been found against security protection mechanisms of WiFi \cite{ccs17} and 4G-LTE systems \cite{Teng13,PETs19}.  
However, all these attacks employ the man-in-the-middle (MITM) attack techniques. 
In the wireless communications, to launch the MITM attacks, an attacker first needs to jam all transmitted signals. 
Thus, the overall timing for communications and cryptographic operations in those wireless systems are important factors to prevent these attacks. 
Another line of research focuses on the security of IoT systems by exploiting networks, e.g., \cite{Sivanathan19,IoT-Traffic,ProfilIoT}.

%
Recently, National Institute of Standards and Technology (NIST) has initiated the lightweight cryptography (LWC) competition to standardize cryptographic algorithm(s) for providing security in resource-constrained environments in the applications of healthcare, Internet of Things, cyber physical systems, and distributed control systems \cite{nist-r2}. 
As a response to the call for proposals, there are 56 submissions received as round 1 candidates in February 2019, and out of 56 candidates, 32 candidates are selected as round 2 candidates in August 2019 \cite{nist-r2}. 
Investigating the performance of such lightweight cryptographic algorithms as cipher suits in different IoT protocols such as WiFi and CoAP running on different IoT devices are important to understand the suitability of such ciphers in the mutual authentication protocols. 

Microcontrollers are a key computing element in IoT in which devices and sensors are equipped with microcontrollers with limited memory, power and processing speed. 
Microcontrollers essentially perform all computations including the security algorithms. 
%
%
As commercial WiFi-enabled devices do not allow developers to implement new algorithms using their APIs, we are unable to leverage existing communication protocols due to the closed platform. 
One way of implementing an OFDM communication system is using software defined radio \cite{Bloessl18,arslan07}.
In this work, we implement the IEEE 802.11a OFDM communication  protocol using the GNU radio and USRPs and cryptographic algorithms in microcontrollers to measure the performance of new lightweight cryptographic algorithms.

\noindent{\bf Our contributions.} 
The goal of this paper is to investigate the performance of three NIST LWC round 2 candidates namely {\ace}, {\spix} and {\wage} in the \textsf{WiFi} and \textsf{CoAP} handshaking mutual authentication and key agreement protocols. 
%
%
Our contributions in this paper are as follows.
\begin{enumerate}[(a)]
\item {\bf Construction and implementation of KDF and MIC.} We first present a construction of a key derivation function (\texttt{KDF}) and  a message integrity check (\texttt{MIC}) generation function relying on a sponge-based authentication scheme as the \texttt{KDF} and \texttt{MIC} are at the core of the IEEE 802.11X and CoAP handshake mutual authentication and key establishment protocols. 
Since a sponge-based AE employs a permutation, our constructions for \texttt{KDF} and \texttt{MIC} involves a minimum invocation of the permutation for generating session keys and {\mic}s. 
We provide optimized implementations of three LWC schemes, namely {\ace} \cite{Aagaard19ace}, {\spix} \cite{Altawy19} and {\wage} \cite{wage_nist_r2} on three different (low-power) microcontrollers (8/16/32-bits) including the \texttt{KDF} and \texttt{MIC} functionalities.  
Our implementations are written in the assembly language and exploit microcontroller resources to achieve a better efficiency. 



\item {\bf Implementing WiFi and the CoAP handshaking authentication in SDR.} We implement the WiFi transportation layer security protocol and the CoAP UDP security protocol using three LWC ciphers, and the IEEE802.11a physical layer communication protocol in SDR in real-time, and embed these two security protocols into the IEEE802.11a communication protocol. The OFDM spectrum in the IEEE802.11a physical layer communication follows the standard from \cite{Keysight_802.11}. Two Universal Software Radio Peripheral (USRP) devices from National Instrument have been used as a sender and a receiver, and the transmission time is captured on the GNU radio running on a PC. 

\item {\bf Experimental evaluation and comparison.} Our experimental setup combines the implementations of SDR and microcontroller codes to obtain the results for handshaking mutual authentication and data protection protocols. 
The execution times of {\kdf} and {\mic} on microcontrollers are measured in Atmel Studio 7.0 and two IAR embedded workbenches which are for MSP430 and Cortex-M3.
We benchmark the performances of three core permutations and present the results for handshaking and data protection protocols on all three microcontrollers. 
Our experimental results show that \ace, {\spix} and {\wage} take about 2,966 ms, 2,831 ms, and 2,808 ms, respectively to complete the IEEE802.11X authentication protocol using \textsf{Cortex-M3}. 
In the data protection protocol, {\ace}, {\spix} and {\wage} achieve a throughput of 109 Kbits/s, 63 Kbits/s, and 53 Kbits/s, respectively on \textsf{Cortex-M3} to encrypt and authenticate a plaintext of 1024 bits and an associated data of 128-bits. 
As the frequency for USRPs used in our experiment is much lower than the actual WiFi systems, our experimental results after scaling to the WiFi frequency show that the cryptographic operations are the dominating factors for authentication and data protection protocols.  

\end{enumerate}
			
 The rest of the paper is organized as follows. Section \ref{sec-background} introduces three LWC schemes, namely {\ace}, {\spix} and {\wage}, the  IEEE802.11X and CoAP handshake protocols, the IEEE802.11a physical layer communication standard, and software defined radio. 
 Section~\ref{sec-Implementation} presents constructions of {\kdf} and {\mic} and their optimized microcontroller implementations including {\ace}, {\spix} and {\wage}.
 In Section \ref{sec-Implementation}, we present the implementation and experiment setup for the handshake and mutual authentication ptotocols in SDR. 
 Section \ref{sec-Results} reports the experiment results of the handshake and data protection protocols and comparisons. Section \ref{sec-concl} concludes the paper.

\section{Preliminaries}\label{sec-background}
In this section, we provide brief backgrounds on three lightweight authenticated encryption schemes, IEEE 802.11X handshake and data protection protocols, CoAP protocol, IEEE 802.11a OFDM standard, and software defined radio.

\subsection{Three Lightweight Authenticated Encryption Schemes}
We consider three lightweight authenticated encryption with associated data (AEAD) schemes, namely {\ace}, {\spix} and {\wage}, which are round 2 candidates in the NIST lightweight cryptography competition \cite{nist-r2}.
\begin{itemize}
\item {\ace} is a lightweight AEAD and hash scheme which operates in the unified sponge duplex mode~\cite{Aagaard19ace} to offer both functionalities.  
At the core of {\ace} is a lightweight permutation of width $320$ bits built upon bitwise XORs and ANDs, left cyclic shifts and 64-bit word shuffles. {\ace} provides a 128-bit security for both AE and hash functionalities.   
\item {\spix} is a lightweight AEAD scheme  which operates in the unified sponge duplex mode~\cite{Altawy19} built upon the \textsf{sLiSCP-light} permutation of width $256$ bits \cite{sliscp-light}. It offers a security level of 128 bits. 
\item  {\wage} is a lightweight AEAD scheme which also operates in  the unified sponge duplex mode \cite{wage_nist_r2}. 
The construction of the {\wage} permutation is based on a Galois-style nonlinear feedback shift register (NLFSR) over the finite field $\F_{2^7}$. 
It accepts a key and a nonce of size 128 bits and  offers a 128-bit security. 
\end{itemize}

For the details about the ciphers and their modes, the reader is referred to \cite{Aagaard19ace,Altawy19,wage_nist_r2}. Table~\ref{tab:param} lists the parameters for AEAD for these three schemes.  The length of each parameter is given in bits and $d$ denotes the amount of processed data (including both associated data (AD), for authentication only,  and message (M) for both encryption and authentication) before a re-keying is done. 
$n$ denotes the internal state size of the permutation, $k$ denotes the key size,  $r$ denotes the rate in the sponge mode, and $t$ denotes the size of the authentication tag.
For each execution of an AEAD algorithm, it processes $r\ell_{AD}$ bits of AD data  and $r\ell_{M}$ bits of plaintext (the padding is applied if AD/M is not a multiple of $r$).

\begin{table}
\begin{center}
\caption{Parameters for {\ace}, {\spix} and {\wage}}
\label{tab:param}
\begin{tabular}{||c|c|c|c|c|c||}
		\hline
	\mbox{Algorithm} & \mbox{State } &
	\mbox{Rate } &
\mbox{Key } &
\mbox{Tag } &
\mbox{Data } \\& $n$ & $r$ & $k$ & $t$ & $\log_2(d)$ \\	\hline 	\hline
\ace & 320 & 64 & 128 & 128 & 124\\	\hline
\spix & 256 & 64 & 128 & 128 & 60\\	\hline
\wage & 259 &  64 & 128 & 128 & 60\\		\hline
\end{tabular}
\end{center}
\end{table}

\subsection{IEEE 802.11i: IEEE 802.11X 4-way handshake and data protection}
In IEEE 802.11X, it is specified that the wireless network consists of supplicants (clients) which wish to be connected to the network, and an access point (or authenticator/server) in the IEEE 802.11X and extensible authentication protocol (EAP) \cite{IEEE8021X,EAP}.
The supplicant and the access point share a pairwise master key ($PMK$) ahead of time. The IEEE 802.11  security solution is specified in the IEEE802.11i amendment.   
To join a network, the device or supplicant executes the 4-way handshake protocol with the authenticator to establish a fresh session key, followed by installing the key (see Figure~\ref{4-way}). 
Once the key is installed, it is used to encrypt  and authenticate traffic data frames using the data protection algorithm. These two phases are summarized as follows: 
 \begin{enumerate}[(a)]
 \item {\bf 4-way handshake protocol}: This process conducts a mutual entity authentication and generates the session keys.
  The 4-way handshake protocol first generates a pairwise transient key ($PTK$) from the pre-shared pairwise master key $PMK$, and then conducts  a challenge-response protocol for mutual authentication.  
 Figure~\ref{4-way} shows an overview of the messages flow in the protocol, where only security related data fields are described. 
 
 \item {\bf Data protection}: After a successful execution of the 4-way handshake protocol, the data protection is performed using either CCMP (AES in counter mode for encryption and CBC MAC for integrity check and message authentication) or  GCMP (AES in counter mode for encryption and polynomial hash for generating a MAC).
\end{enumerate}
The WiFi data field contains identifiers, key information, replay counter, nonce, initial vector, message integrity code (MIC), and transported data. 
In Figure~\ref{4-way}, for simplicity, we omit the format of EAP and the case using group keys. We show only cryptographic functionalities involved in the 4-way handshake protocol. 
\begin{figure}[ht]
\begin{center}
\begin{tikzpicture} [rounded corners=2pt]
\begin{scope}[>=latex]

\draw (0, 2.75) node[anchor=south] {\small{Supplicant}};
\draw (0, 3.0) node[minimum height=0.6cm,minimum width=2.6cm, draw] {};

\draw (4, 2.8) node[anchor=south] {\small{Authenticator}};
\draw (4, 3.0) node[minimum height=0.6cm,minimum width=2.6cm, draw] {};

\draw (0, 2.7) -- (0, -1.8);
\draw (4, 2.7) -- (4, -1.8);

\draw (2, 1.8) node[anchor = south] {\scriptsize{$ANonce$}};
\draw[<-] (0, 1.8) -- (4, 1.8);

\draw (5.3, 1.9) node[anchor = south] {\scriptsize{Generate $ANonce$}};

\draw (2, 0.9) node[anchor = south] {\scriptsize{($SNonce$, $MIC_A$)}};

\draw[->] (0, 0.9) -- (4, 0.9);

\draw (-1.2, 1.0) node[anchor = south] {\scriptsize{Generate $SNonce$}};
\draw (-1.0, 0.7) node[anchor = south] {\scriptsize{Derive $PTK$}};

\draw (2, 0) node[anchor = south] {\scriptsize{($ANonce$,  $MIC_{S}$)}};
\draw[<-] (0, 0) -- (4, 0);

\draw (5, 0.5) node[anchor = south] {\scriptsize{Derive $PTK$}};

\draw (2, -0.9) node[anchor = south] {\scriptsize{$MIC_{all}$}};
\draw[->] (0, -0.9) -- (4, -0.9);

\draw (5, -1.1) node[anchor = south] {\scriptsize{Install $TK$}};
\draw (-0.8, -1.1) node[anchor = south] {\scriptsize{Install $TK$}};

\end{scope}

\end{tikzpicture}
\caption{Message flow of the IEEE 802.11X 4-way handshake protocol}\label{4-way}
\end{center}
\end{figure}

\noindent{\bf PKT and MIC generation in 4-way handshake.} 
The pairwise transit key ($PTK$) is generated as follows
\begin{align}
PTK \nonumber = & \texttt{KDF}(PMK,  ANonce || SNonce ||   \mbox{AP\,MAC\,adr} ||  \\ & \nonumber \mbox{STA\,MAC\,adr}) = KCK||KEK||TK
\end{align}
where \texttt{KDF} is a key derivation function,  the nonces namely $ANonce$ and $SNonce$ are 128 bits. 
The first 128-bit in $PTK$ is the key confirmation key ($KCK$) that is used to generate a MIC over the message, the second 128-bit is the key encryption key ($KEK$) that is used for encrypting the group key, and the last segment is the temporal key ($TK$) used for protecting traffic data where the length depends on a cipher suite selected. 
\begin{equation}
	\begin{array}
		{l}
		MIC_{A}= \mic (KCK,  ANonce, RC)\\
		MIC_{S} = \mic(KCK,  SNonce, RC)\\
		MIC_{all}= \mic(KCK, D, RC+1)
	\end{array}
\end{equation}
where $RC$ is a replay counter of 128 bits (see \cite{Chen-Gong-book12}), and $D$ carries  the cipher suite of 128 bits.

\subsection{CoAP: DTLS Handshake and Data Protection Protocols}
The Constrained Application Protocol (CoAP) enables an efficient transmission of information for resource-limited devices \cite{rfc7252}. 
The security in CoAP is provided using Datagram TLS (DTLS) over the user datagram protocol (UDP). 
Figure \ref{6-way} shows the message flow of the protocol when a server authenticates an IoT client device in an IoT network. 
There are four security modes available in CoAP, namely  \textsf{NoSec}, \textsf{PresharedKey}, \textsf{RawPublicKey} and \textsf{Certificates}  \cite{rfc7252,Rahman16}. 
\textsf{NoSec} means the security is not provided in the CoAP message transmission. 
\textsf{PresharedKey} mode is used for symmetric-key algorithms for authentication and message protection. 
\textsf{RawPublicKey} mode is used for the public-key algorithms without certificate, and the devices are programmed with a list of pre-installed keys. 
\textsf{Certificates} mode provides authentication based on the X.509 public-key certificate.

Note that both \textsf{Certificates} and \textsf{RawPublicKey} use elliptic curve (EC) based public key cryptography, and 
\textsf{PresharedKey} uses TLS-PSK based on symmetric-key algorithms with the cipher suit TLS-PSK-WITH-AES128-CCM-8 for authentication. 
\textsf{Certificates} mode uses the cipher suit TLS-ECDHE-ECDSA-WITH-AES-128-CCM-8 with X.509 certificate. 
\textsf{RawPublicKey} mode uses the cipher suit TLS-ECDHE-ECDSA-WITH-AES-128-CCM-8. 

In our work, we consider only the \textsf{PresharedKey} mode for authentication. Note that in Figure \ref{6-way}, $ClientHello$ contains the client's version number ($ver._C$), client random nonce ($Nonce_C$), session ID ($ID_C$), cipher suit ($Ciphersuit_C$) and compression method ($Compress_C$) \cite{Hamdane11}. 
Similarly, for the server, $ServerHello$ contains the server's nonce ($Nonce_S$), session ID ($ID_S$), cipher suit ($Ciphersuit_S$), and compression method ($Compress_S$).
Like IEEE 802.11X, the client and server share a pairwise master key (PMK). 
To joint the network, the device executes the 6-way handshaking with the server to establish a fresh session key, followed by installing the key (TK) after the 6-th round of the handshake.  
After that, the device uses the installed key to encrypt and authenticate traffic data frames using the protection algorithms.
We summarize these two phases below:
 \begin{enumerate}[(a)]
 \item {\bf CoAP 6-way handshake protocol:} This conducts the mutual entity authentication and generation of session keys, as shown in Figure~\ref{6-way}.
 \item {\bf Data protection:} After a successfully completing the handshake protocol, the data protection algorithm, which is AES128 in counter mode with CBC-MAC and 8-octet Integrity Check Value (ICV), is applied to secure the traffic.
  \end{enumerate}
%
Similar to IEEE 802.11X, the 6-way handshake generates a pairwise transient key (PTK) from $PMK$, and conducts  a challenge-response protocol for mutual authentication.
\begin{figure}[ht]
\begin{center}
\resizebox{8.5cm}{!}{
\begin{tikzpicture} [rounded corners=2pt] 
\begin{scope}[>=latex]

\draw (0, 2.75) node[anchor=south] {\small{Client}};
\draw (0, 3.0) node[minimum height=0.6cm,minimum width=2.6cm, draw] {};

\draw (4, 2.75) node[anchor=south] {\small{Server}};
\draw (4, 3.0) node[minimum height=0.6cm,minimum width=2.6cm, draw] {};

\draw (0, 2.7) -- (0, -3.3);
\draw (4, 2.7) -- (4, -3.3);

\draw[->] (0, 1.8) -- (4, 1.8);

\draw (-1.6, 1.8) node[anchor = south] {\scriptsize{Generate $ClientHello$}};

\draw (2, 2.05) node[anchor = south] {\scriptsize{$ClientHello=Ver_C$,$Nonce_C$,}};

\draw (2, 1.75) node[anchor = south] {\scriptsize{$ID_C$, $Ciphersuit_C$,$Compress_C$}};

\draw[<-] (0, 0.9) -- (4, 0.9);

\draw (2, 0.9) node[anchor = south] {\scriptsize{$MAC(ClientHello)$}};

\draw (5, 1.1) node[anchor = south] {\scriptsize{Generate}};

\draw (5.3, 0.8) node[anchor = south] {\scriptsize{HelloVerifyRequest}};
\draw (5.3, 0.5) node[anchor = south] {\scriptsize{(contains cookie)}};

\draw (2, 0.3) node[anchor = south] {\scriptsize{$ClientHello$}};

\draw (2, 0) node[anchor = south] {\scriptsize{$MAC(ClientHello)$}};

\draw (5.3, -0.5) node[anchor = south] {\scriptsize{Derive $MIC_C$}};

\draw[->] (0, 0) -- (4, 0);

\draw (5.6, -0.8) node[anchor = south] {\scriptsize{Generate $ServerHello$}};
\draw (5.3, -1.2) node[anchor = south] {\scriptsize{Generate $PTK$}};

\draw (2, -0.6) node[anchor = south] {\scriptsize{$MIC_C$}};

\draw (2, -0.9) node[anchor = south] {\scriptsize{$ServerHello=$$Nonce_S$,$ID_S$,}};

\draw (2, -1.2) node[anchor = south] {\scriptsize{$Ciphersuit_S$,$Compress_S$}};

\draw[<-] (0, -1.1) -- (4, -1.1);
\draw (-1.15, -1.8) node[anchor = south] {\scriptsize{Derive $MIC_S$}};
\draw (-1.15, -1.4) node[anchor = south] {\scriptsize{Derive $PTK$}};

\draw (2, -2.1) node[anchor = south] {\scriptsize{$ServerHello$, $MIC_S$}};

\draw[->] (0, -2) -- (4, -2);

\draw (2, -2.95) node[anchor = south] {\scriptsize{$MIC_{all}$}};

\draw (5.3, -2.7) node[anchor = south] {\scriptsize{Generate $MIC_{all}$}};

\draw[<-] (0, -2.9) -- (4, -2.9);

\draw (-0.8, -3.3) node[anchor = south] {\scriptsize{Install TK}};
\draw (4.8, -3.3) node[anchor = south] {\scriptsize{Install TK}};


\end{scope}

\end{tikzpicture}}

\caption{Message flow of the 6-way CoAP handshake protocol}\label{6-way}
\end{center}
\end{figure}
The PTK and MIC generation are similar to that of IEEE 802.11X, except the following 
\begin{align}
PTK & \nonumber = \texttt{KDF}(PMK,  ClientHello || ServerHello) \\
    & \nonumber = KCK||KEK||TK
\end{align}
where {\kdf} is a key derivation function. Here we omit the {\mic} generations for CoAP. 

\subsection{IEEE 802.11a OFDM Standard}\label{subsec-IEEE_802.11_Standard}
\begin{figure}[h]
  \centering
  \includegraphics[width=\linewidth]{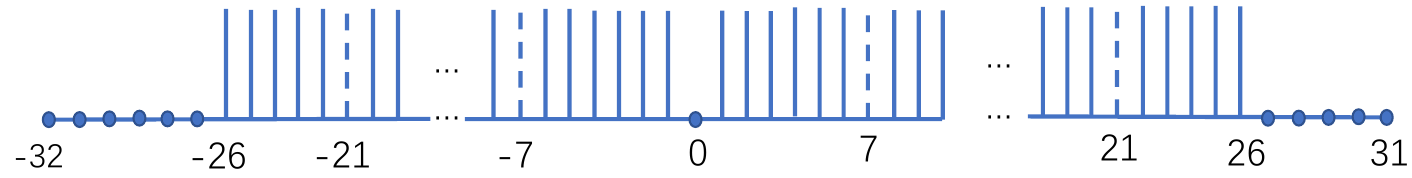}
  \caption{802.11a OFDM spectrum 64 subcarriers usage}\label{fig:OFDM_802.11a_Keysight}
\end{figure}

We now introduce the IEEE 802.11a physical layer orthogonal frequency-division multiplexing (OFDM) system. 
See Appendix~\ref{sec:OFDM} for the general definition of the OFDM system. 
Figure \ref{fig:OFDM_802.11a_Keysight} shows the channel usage for the 64-IFFT OFDM system in the IEEE 802.11a \cite{Keysight_802.11}. 
In this standard, the subcarriers are labeled from $-32$ to $31$ and only 48 subcarriers are used to transmit data and 4 subcarriers $(-21, -7, 7, 21)$ 
are used for pilot carriers for the channel estimation. 
The unused subcarriers from -31 to -25 and from 27 to 31 are used to reduce the leakage of 52 subcarriers' sidelobes power to the outside of the total bandwidth. 
Finally, the direct current (DC) is labeled as 0 subcarrier. Note that the IEEE 802.11a OFDM system does not use the DC subcarrier to transmit the information;
Therefore, the DC subcarrier is inserted a complex number 0 at the carrier allocator before the inverse fast fourier transformation (IFFT) is done.

In summary, the IEEE 802.11a OFDM system has the following specifications.
\begin{itemize}
  \item The total bandwidth is $20$ MHz, and the total number of subcarriers is $52$ from $-26$ to $26$ (not include DC at 0) where the subcarriers from -32 to -27 and from 27 to 31 are not used. 
  \item The underlying modulation could be BPSK, QPSK, 16-QAM, and 64-QAM.
  \item There are $48$ data subcarrier and $4$ pilot subcarriers used for the channel estimation. The pilot subcarriers are located at $-21, -7,7,21$, and pilot symbols are modulated by BPSK.
  \item The information rate could be $6, 9, 12, 18, 24, 36, 48,$ and $54$ Mbits/s. 
\end{itemize}

\subsection{Software Defined Radio}
\begin{figure}
  \centering
  \includegraphics[width=\linewidth]{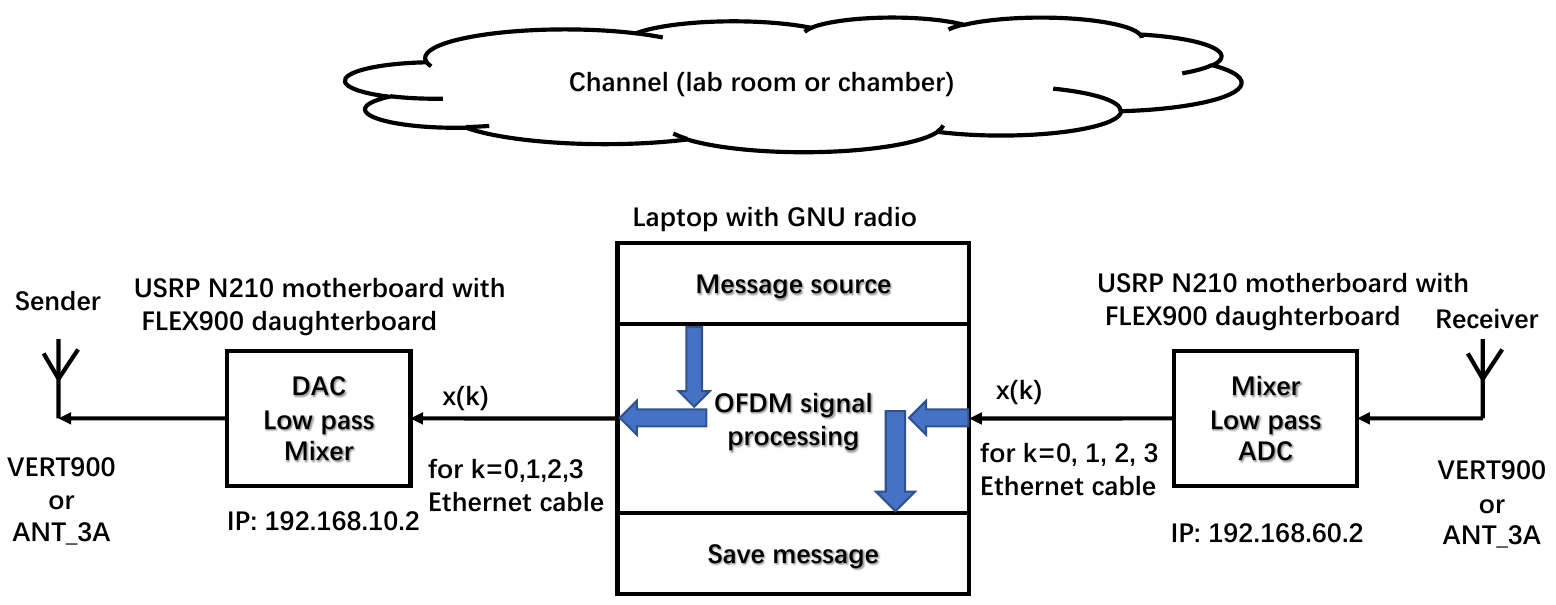}
  \caption{Software defined radio physical connection}\label{fig:SDR_connection}
  \vspace{-0.1in}
\end{figure}

Figure~\ref{fig:SDR_connection} provides a high-level overview of our setup of Software Defined Radio (SDR). 
The SDR consists of two main parts, namely universal software radio peripheral (USRP) and GNU radio. 
The USRP is a hardware composed of ADC, DAC, low pass filer and mixer. 
The GNU radio is a digital signal processing (DSP) software running on a Linux operating system (PC). 
By setting up an IP address to a USRP block in the GNU radio, the GNU radio on a PC can send and receive sample data from the USRP hardware devices. 
An ethernet cable physically creates a connection between a USRP device and the GNU radio. 
Since each USRP device has its own subnet, the PC uses two network cards to connect two USRP devices simultaneously. 
Instead of using a gigabit ethernet switch \cite{Tran14}, we use two gigabit ethernet adaptors to connect two USRPs separately and set up their own IP addresses and subnets. 
After running the SDR, the data transmission among USRP devices and the GNU radio are recorded in real-time. 
As a result, we can use the QT-GUI-frequency-sink block in the GNU radio to analyze the spectrum of sampled data obtained from the USRP devices in real-time.

The GNU radio is a software working under the GUN General Public License which is a copyleft license and free to use \cite{Yunjie18}. 
The GNU radio companion is a graphical software combining those functions to form graphical intergraded blocks, which helps to develop prototyping and signal processing graphically. 
In our experiment, we use a GNU radio companion Ver. $3.7.9$ on Ubuntu $16.04$ OS which is further running on a visual machine on a Windows 10 machine.

The GNU radio on a PC is used to process digital signals in a communication system, and the connected USRP devices process the analog signals. 
For the sender, the ethernet cable is used to transfer the discrete time sample data from the GNU radio to the USRP device (left side in Figure \ref{fig:SDR_connection}) and 
then the USRP device converts it into analog signals and passes them through the low pass filter, the mixer, and the antenna. 
For the receiver, the USRP device (right side in Figure \ref{fig:SDR_connection}) receives the analog signal from its antenna and converts it to baseband signals. 
Finally, the filtered signals are converted into discrete time signals, and then send to the PC through the ethernet cable.
There are five main data types available in the GNU radio which are byte (8 bits), short (16 bits), integer (32 bits), float (32 bits), and complex (two 32-bit floats).

\section{Construction and Implementation of \texttt{KDF} and \texttt{MIC} in CoAP and IEEE 802.11i}\label{sec-Implementation}
\subsection{\ace, \spix, and {\wage} as cipher suites in CoAP and IEEE 802.11i Protocols}\label{sec:Const_MIC_KDF}
In IEEE 802.11X 4-way and CoAP $6$-way handshake protocols, the key derivation function (\kdf) and message integrity check (\mic) are two fundamental cryptographic functionalities for authentication.  
The data protection protocol requires an AEAD algorithm for encryption and tag generation to protect the traffic.  
Our idea is to use a single cryptographic primitive (e.g., AEAD) to serve all cryptographic functionalities required in the handshake and data protection protocols. 
We now show how to construct {\kdf} and {\mic} algorithms from the \ace, \spix, and {\wage} AEAD schemes. 
Let $\Fn$ be the underlying permutation instantiating an AEAD scheme where $\Fn \in\{\ace,\spix,\wage\}$.
As all three ciphers operate in the sponge duplex mode \cite{duplexing,sliscpsac2017} with different permutations of different widths and the rate for absorbing messages are the same, 
we provide a generic construction that works for $\Fn \in \{\ace,\spix,\wage\}$. 

\begin{figure*}[h]
  \centering
  \includegraphics[width=\linewidth]{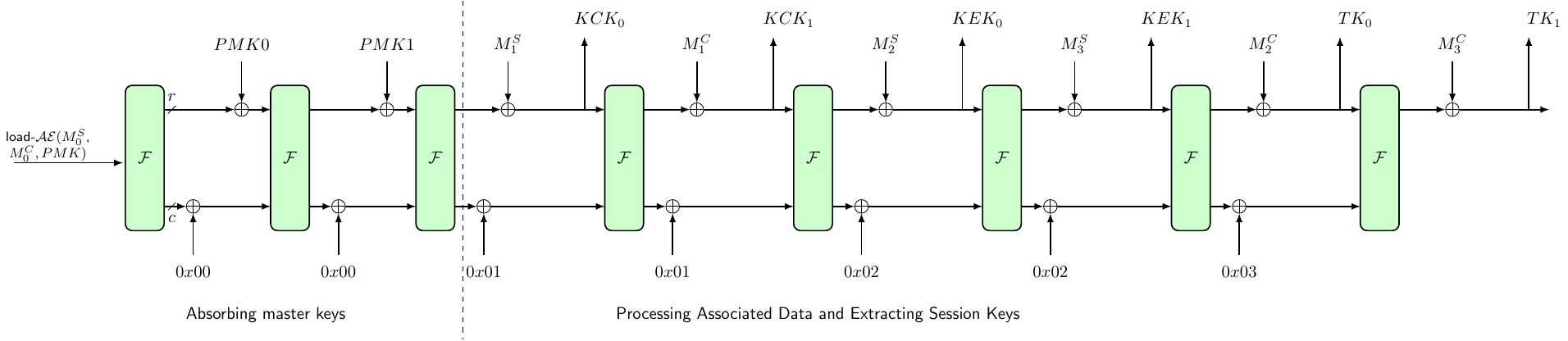}
  \caption{A diagram of the key derivation function (KDF) deriving $KCK$, $KCK$ and $TK$.}\label{fig:KDF}
\end{figure*}

\noindent{\bf Constructing \kdf.} 
We now show how to configure a sponge-based AEAD as a key derivation function in both handshake protocols. 
As an AEAD scheme has three phases, we use only the initialization and encryption phases to construct a {\kdf}, 
but with a subtle difference that for each key type $KCK$, $KEK$ or $TK$,  we use different domain separation values.
\begin{construction}[Key Derivation Function]\label{const:KDF}
 Let $S = (S_r,S_c)$ be the state of the permutation $\Fn \in\{\ace,\spix,\wage\}$ where $S_r$ with $|S_r| = r$ and $S_c$ with $|S_c| = c$ denote the rate part and capacity part of the state, respectively, and  $n = r + c$ denotes the state size. Let $M^C$ and $M^S$ be the inputs of length 256 bits from the client (or supplicant) and server (or access point) in {\kdf} to derive session keys using $PMK$.  
 Let $M^C = M_0^C \| M_1^C \| M_2^C \| M_3^C$ and $M^S = M_0^S \| M_1^S \| M_2^S \| M_3^S$ where $|M_i^C| = |M_i^S| = r = 64$. 
 Then the key derivation function to derive $PTK = KCK \| KEK \| TK$ is defined as 
 \begin{itemize}
  \item[-] {\bf Loading master key:} $$S \leftarrow \textsf{load}(M_0^S\|M_0^C, PMK); S \leftarrow \Fn(S)$$
  \item[-]{\bf Absorbing key:}
  \begin{align}
  & S \nonumber \leftarrow (S_r\oplus PMK_0, S_c ); S \leftarrow \Fn(S) \\
  & S \nonumber \leftarrow (S_r\oplus PMK_1, S_c );  S \leftarrow \Fn(S)
 \end{align}
 \item[-] {\bf Outputting $KCK = KCK_0\|KCK_1$:}
  \begin{align}
  & KCK_0 \nonumber \leftarrow S_r\oplus M^S_1;  \\ &  S \nonumber \leftarrow (KCK_0, S_c\oplus(0^{c-2}\|01)); S \leftarrow \Fn(S)\\
  & KCK_1 \nonumber \leftarrow S_r\oplus M^C_1; \\ & S \nonumber \leftarrow ( KCK_1, S_c\oplus(0^{c-2}\|01));  S \leftarrow \Fn(S)
 \end{align}
 \item[-] {\bf Outputting $KEK= KEK_0\|KEK_1$:}
 \begin{align}
  & KEK_0  \nonumber \leftarrow S_r\oplus M^S_2; \\ & \nonumber  S \nonumber \leftarrow (KEK_0, S_c\oplus(0^{c-2}\|10)); S \leftarrow \Fn(S) \\
  & KEK_1  \nonumber \leftarrow S_r\oplus M^S_3; \\ &  S \nonumber \leftarrow ( KEK_1, S_c\oplus(0^{c-2}\|10));  \nonumber  S \leftarrow \Fn(S)
 \end{align}
  \item[-] {\bf Outputting $TK= TK_0\|TK_1$:}
 \begin{align}
  & TK_0  \leftarrow S_r\oplus M^C_2; S \nonumber \leftarrow (TK_0, S_c\oplus(0^{c-2}\|11));  \\
  & S \leftarrow \Fn(S); TK_1 \nonumber \leftarrow S_r\oplus M^C_3. 
 \end{align}
 \end{itemize}
\end{construction}
We provide a pictorial representation of the {\kdf} in Figure~\ref{fig:KDF}. Note that the size of the capacity part depends on $\Fn$. 
In practice, the MAC address of a device is a 48-bit number. However, we convert it into a 128-bit number by applying the padding 1 followed by 79 zeros (i.e., $10^{79}$). 
The reason for making the length of MAC addresses is due to the output length of $PTK$.

\begin{figure*}[h]
  \centering
  \includegraphics[width=0.9\linewidth]{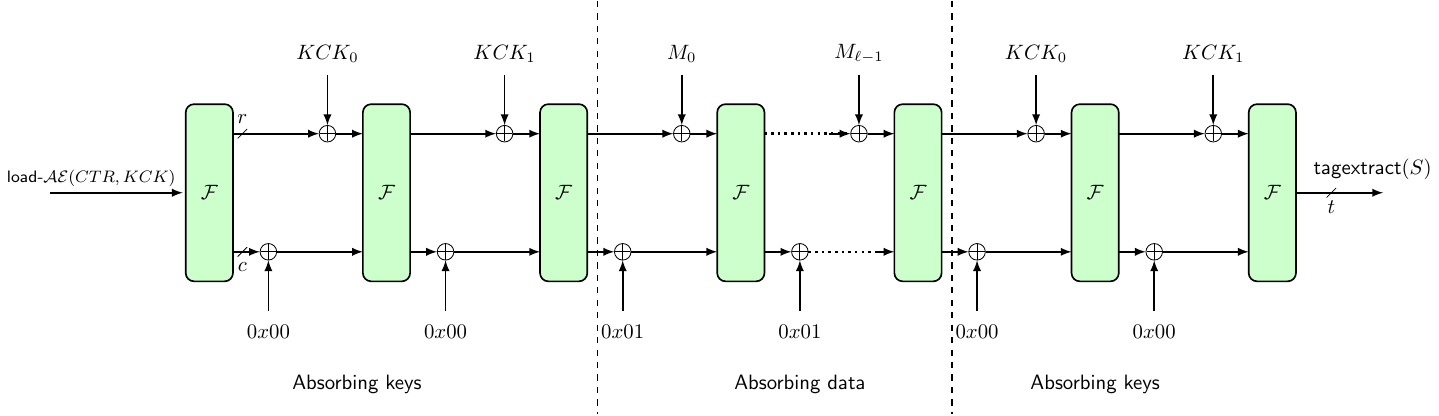}
  \caption{A block diagram of the {\mic} generation function. \textsf{tagextract}$(S)$ extracts the tag from pre-defined positions in $S$.}\label{fig:MIC}
\end{figure*}

\noindent{\bf Constructing \mic.} 
Our idea for adopting a sponge-based AEAD scheme to construct a {\mic} is by computing a tag on a non-empty associated data and an empty plaintext (with no padding). 
Note that in the IEEE 802.11X 4-way and CoAP 6-way handshake protocols, the session key $KCK$ is used to generate three MICs on $ANonce (ServerHello)$, $SNonce (ClientHello)$ and $D$ fields. 
Below we provide a construction of a {\mic} based on an AEAD scheme and present a pictorial representation of it in Figure \ref{fig:MIC}. 
\begin{construction}[Message Integrity Code]\label{const:MIC}
Let $CTR = CTR_0 \| CTR_1$ be a counter and $M = M_0 \| \cdots \|M_{\ell-1}$ be a message of $\ell$ blocks after padding. 
Following the notations in Construction~\ref{const:KDF}, the message integrity code on $M$ and $CTR$ is constructed as follows. 
 \begin{itemize}
  \item[-]{\bf Loading and absorbing KCK:}
  \begin{align}
  & S \nonumber \leftarrow \textsf{load}(CTR, KCK); S \leftarrow \Fn(S)\\
  & S \nonumber \leftarrow (S_r\oplus KCK_0, S_c ); S \leftarrow \Fn(S) \\
  & S \nonumber \leftarrow (S_r\oplus KCK_1, S_c );  S \leftarrow \Fn(S)
 \end{align}
 \item[-] {\bf Absorbing $M$:} $i = 0 \cdots \ell-1$
  \begin{align}
  & S \nonumber \leftarrow (S_r \oplus M_{i}, S_c\oplus(0^{c-2}\|01)); S \leftarrow \Fn(S)
 \end{align}
 \item[-] {\bf Absorbing KCK again:} 
 \begin{align}
  &  S \nonumber \leftarrow (S_r\oplus KCK_0, S_c )); S \leftarrow \Fn(S) \\
  &  S \nonumber \leftarrow ( S_r\oplus KCK_1, S_c);  \nonumber  S \leftarrow \Fn(S)
 \end{align}
 \item[-] {\bf Outputting MIC:} $MIC \leftarrow \textsf{tagextract}(S)$. 
 \end{itemize}
\end{construction}
For example, in IEEE 802.11X, while generating $MIC_A$, $M = ANonce$ and $CTR = RC$, while generating $MIC_S$, $M = SNonce$ and $CTR = RC$, and while generating $MIC_{all}$, $M = D$ and $CTR = RC+1$ for the above construction. 

\noindent{\bf Security and efficiency.}
Intuitively, the security of Constructions~\ref{const:KDF} and~\ref{const:MIC} relies on the security of the AEAD algorithm. 
Following the parameters and security of {\ace}, {\spix} and {\wage}, the security of both {\kdf} and {\mic} is 128 bits \cite{Aagaard19ace,Altawy19,wage_nist_r2}. 
The efficiency of both {\kdf} and {\mic} is measured by the number of permutation calls required to complete the functionality. 
As the rate $r$ is 64, the {\kdf} in Construction~\ref{const:KDF} needs {\it eight} calls to the permutation as it outputs three session keys and each of length 128 bits. 
On the other hand, the {\mic} in Construction~\ref{const:MIC} needs {\it ($\ell$+5)} calls to the permutation as the initialization and finalization needs five calls and absorbing the message needs $\ell$ calls. 

\vspace{-0.25cm}

\subsection{Microcontroller Implementations of \ace, {\spix} and \wage}\label{subsec-Impl}
This subsection presents the details about the microcontroller platforms and the implementations of three ciphers. 
 
\noindent\textbf{Microcontroller platform.} 
We implement \ace, \spix, and {\wage} and corresponding {\kdf} and {\mic} algorithms described in Section~\ref{sec:Const_MIC_KDF} in assembly on three different microcontrollers, namely 8-bit \textsf{Atmega128}, 16-bit \textsf{MSP430F2013/2370} and 32-bit \textsf{Cortex-M3LM3S9D96}. 
The \textsf{IAR} embedded workbenches for \textsf{MSP430} and \textsf{Cortex-M3} and the \textsf{Atmel Studio} 7.0 for \textsf{Atmega128} have been used to import the codes into microcontrollers and 
to calculate the number of clock cycles and the execution time for the AEAD, {\kdf} and {\mic} algorithms. Table~\ref{tab:specification_mcpu} summarizes the resources such as the flash memory size, the RAM size and the number of general-purpose registers available on three microcontrollers. 
The throughput, denoted by $\eta$,  is calculated as $\eta = \frac{m\times f}{C}$ where $m$ is the length of the message, $f$ is the CPU frequency and $C$ is the total number of clock cycles. 
The CPU frequency for all three microcontrollers used in our experiment is $16$ MHz. 
We report the memory usage, the number of clock cycles from the IAR embedded workbenches and the Atmel Studio in the debug mode. 
\begin{table}[h]
\centering
\caption{Resources of 8/16/32-bit microcontrollers}
\label{tab:specification_mcpu}
\resizebox{8cm}{!}{
\begin{tabular}{||l|c|c|c||}
\hline
{\bf Microcontrollers} & {\bf Flash memory}   & {\bf RAM} {[}kB{]} & {\bf Number of general } \\  
		& size [kB] &  &  {\bf purpose registers} \\ \hline \hline
\textsf{ATmega128}                 & 128                        & 4.448        & 32(R0 - R31)                       \\ \hline
\textsf{MSP430F2013}                & 2.304                      & 0.128        & 12 (R4 - R15)                      \\ \hline
\textsf{MSP430F2370}                     & 33.024                     & 2.048        & 12 (R4 - R15)                      \\ \hline
\textsf{Cortex-M3LM3S9D96}  & 524.288                    & 131.072      & 13 (R0 - R12)                      \\ \hline
\end{tabular}
}
\end{table} 


\noindent\textbf{{\spix} and {\ace}.} 
In our implementation, we target to achieve a highest level of throughput. 
For the 8-bit \textsf{Atmega128} implementation, the state of the {\spix} permutation is stored in registers so that we can avoid data exchange between the memory and registers while executing the round function of the permutation. On the other hand, for the 16-bit and 32-bit microcontroller implementations, the states of  {\spix} and {\ace} are stored into the memory due to not having available registers to entirely store the state. While executing the permutation, 
the partial state is stored into registers and then after partial state update, it is again stored back to the memory.   
The most expensive operation is shifting state words, so that the method above saves clock cycles by managing the position of the state instead of shifting the contents of the state.
Note that the {\ace} permutation requires more registers than that of the {\spix} permutation due to the larger state size.\\*

\noindent\textbf{{\wage}.}
For the 16-bit microcontroller implementation of {\wage}, we use \textsf{MSP430F2370}, instead of \textsf{MSP430F2013}, due to a larger memory space to save the round constants.
The design of the {\wage} permutation is based on a shift register which requires shifting the entire state for each execution of the round function, which consumes 36 shift operations in each iteration. 
Instead of loading the state of 259 bits into registers, the state is contiguously stored in the RAM.  
To execute the permutation, we extract the corresponding 7-bit words from the RAM into registers and apply the permutation operations such as lookup table and bitwise XOR operations. 
After computing the feedback, the updated value is stored next to the memory location of the current state. In this way, for each iteration we allocate a new memory byte, which results in $148\;(=37+111)$ bytes for 111 rounds of the {\wage} permutation. 

The absolute locations of those extracted content in the RAM are not fixed but the relative locations to the first byte of the 259 bits are fixed. Therefore, we only need the initial memory location of the first byte of the 259 bits, which is the same for each round,  denoted as $\textsf{init}$, the integer numbers of the relative locations, denoted as set $\lambda$, and an integer variable $\textsf{init}$ to record the current round number. Then, the current locations of the extracted contents will be the set $\{\textsf{init}  + \textsf{init} +t \,|\, t\in \lambda\}$. After finishing 111 rounds, we set $\textsf{init} = 0$ and copy the final state to the initial state location in the RAM. 
Then, we proceed to the next evaluation of the {\wage} permutation.

\section{Implementation of OFDM System, IEEE 802.11X and CoAP in SDR}\label{subsec-Impl}
\begin{figure}[ht]
  \centering
  \includegraphics[width=0.95\linewidth]{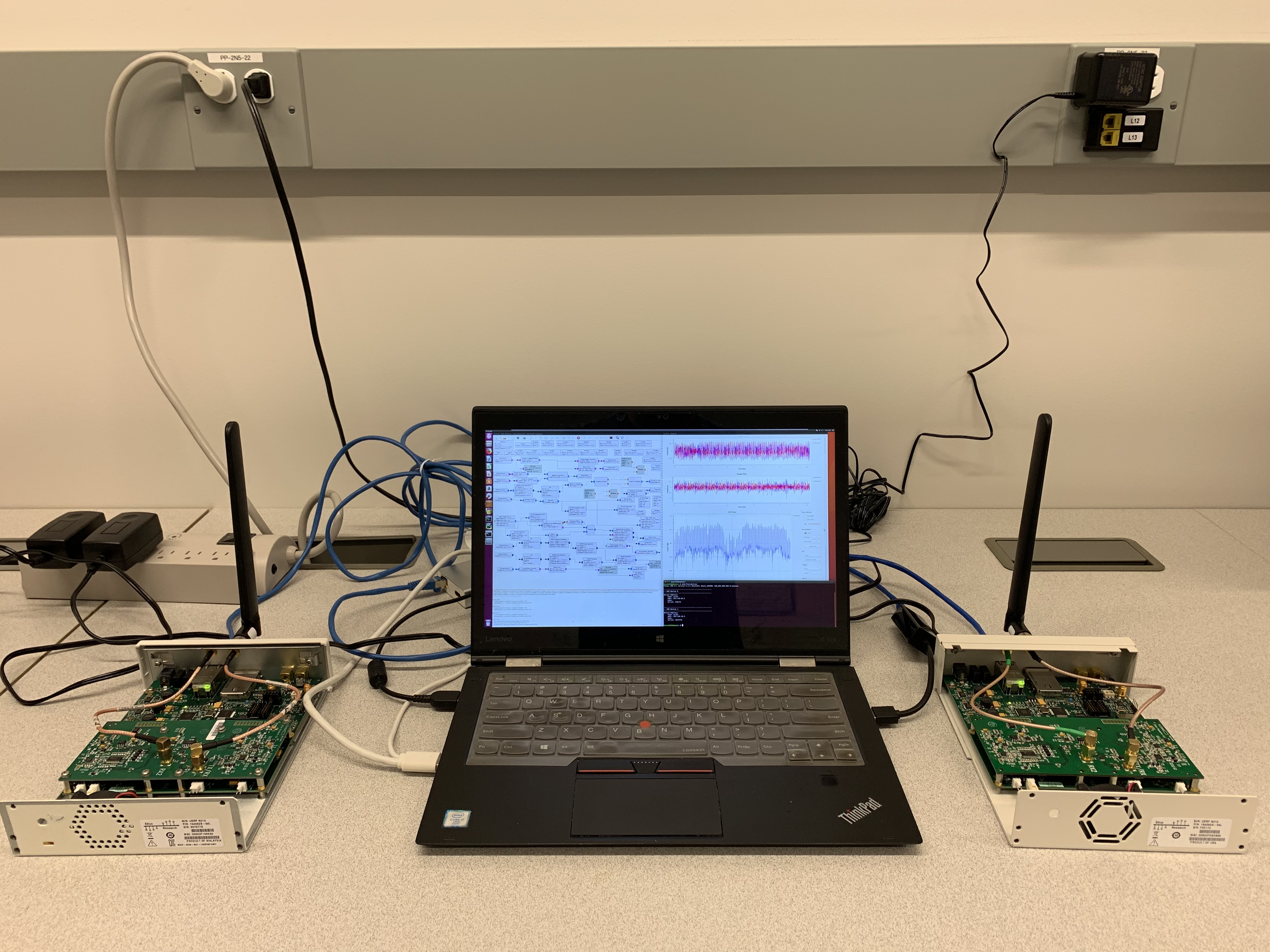}
  \caption{Experimental setup of software defined radio}\label{fig:expsetup}
\end{figure}

We implement the OFDM system in the GNU software defined radio (SDR). In our implementation, the OFDM system consists of an OFDM sender, an OFDM receiver, and the GNU radio companion. 
Figure~\ref{fig:expsetup} provides our experimental setup for the OFDM system.  We now describe the implementation details of the OFDM sender and the OFDM receiver. 

\subsection{Experimental Setup for OFDM Sender and Receiver}
\begin{figure*}[h]
  \centering
  \includegraphics[width=\linewidth]{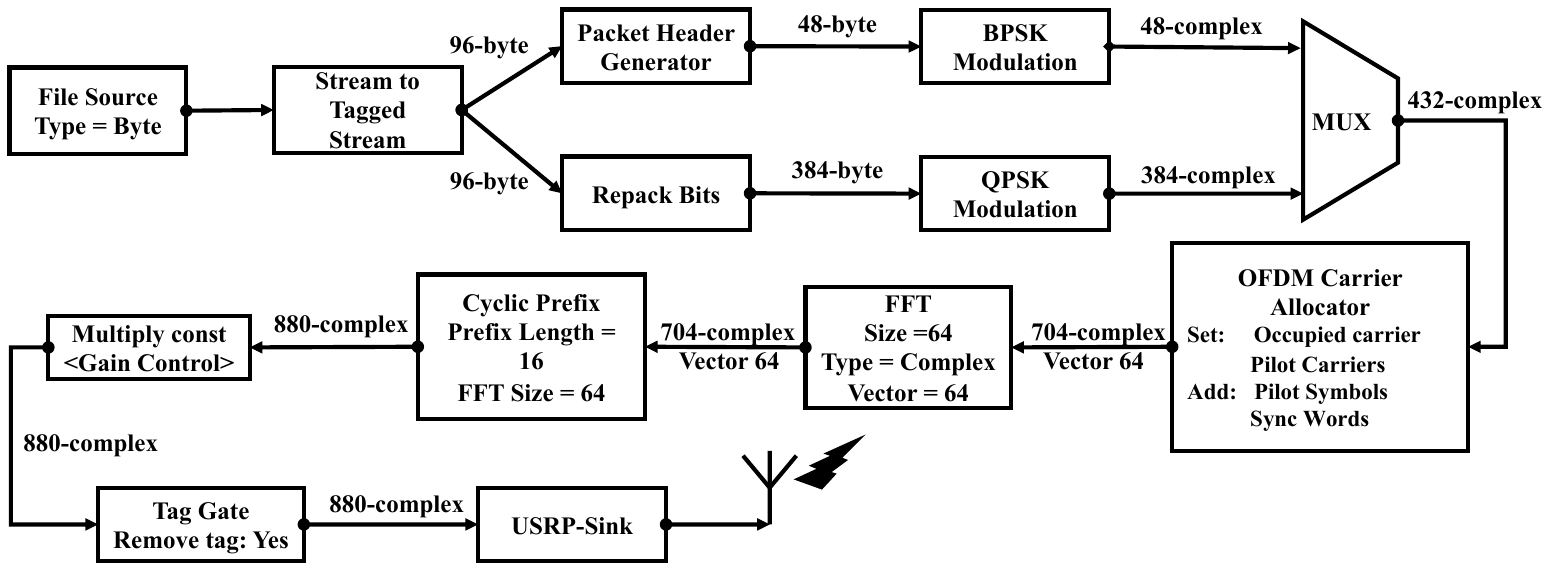}
  \caption{A block diagram of the OFDM sender}\label{fig:OFDM_Sender_without_Enc}
\end{figure*}

\noindent\textbf{IEEE 802.11a OFDM Sender.} Figure~\ref{fig:OFDM_Sender_without_Enc} shows an overview of the OFDM sender. The file-source block in the GNU radio is used to output bytes from a binary file to its next block, and it is set to repeatedly sending the message bit stream automatically during each test. Each $96$ bytes from the file-source block will be tagged in the stream-to-tagged-stream block. After that, the following blocks will manipulate each $96$ message bytes at a time. For example, the packet-header-generator block generates $48$ header bytes for each tagged message which is the tagged 96-byte.

The repack-bits block in Figure \ref{fig:OFDM_Sender_without_Enc} operates 1-byte at a time. We denote a 1-byte input as $\mb{a} = (a_0,a_1,a_2,a_3,a_4,a_5,a_6,a_7).$ 
The repack-bits block converts $\mb{a}$ to $\mb{d}=(d_0,d_1,d_2,d_3)$ by converting each 2-bit $(a_{2i}, a_{2i+1})$ to a decimal number $d_i$. 
After that, each decimal number is converted to a byte $\mb{b}$ which is $\mb{b}=(b_0,b_1,b_2,b_3).$
The output of the repack-bits block is $\mb{b}=(b_3,b_2,b_1,b_0)$ which is in the endianness of LSB.
Comparing the input $\mb{a}$ with the output $\mb{b}$, it indicates that each input byte corresponds to four output bytes, which explains that each 96-byte input has 384 bytes for the repack-bits block in Figure \ref{fig:OFDM_Sender_without_Enc}.

The BPSK-modulation block also converts each byte to an 8-byte complex number. More specifically, it maps the tuple of bytes $(00 ,01)_{hex}$ to complex numbers $((-1, 0), (1, 0))_{decimal}$.
Similarly, the QPSK-modulation block converts each byte input into a complex number. It maps the input bytes $(00 ,01, 02, 03)_{hex}$ to output complex numbers $((-1/\sqrt{2}, -1/\sqrt{2})$, $ (1/\sqrt{2}$, $ -1/\sqrt{2})$, $(-1/\sqrt{2}, 1/\sqrt{2})$, $(1/\sqrt{2}, 1/\sqrt{2}))_{decimal}$, respectively.

The MUX block is used to combine each $48$-complex header and $384$-complex payload at a time. Therefore, the output of the MUX block is $432$ complex numbers in total, and is sent to the OFDM-carrier-allocator block, which is described as follows.

\begin{figure*}[ht]
  \centering
  \includegraphics[width=\linewidth]{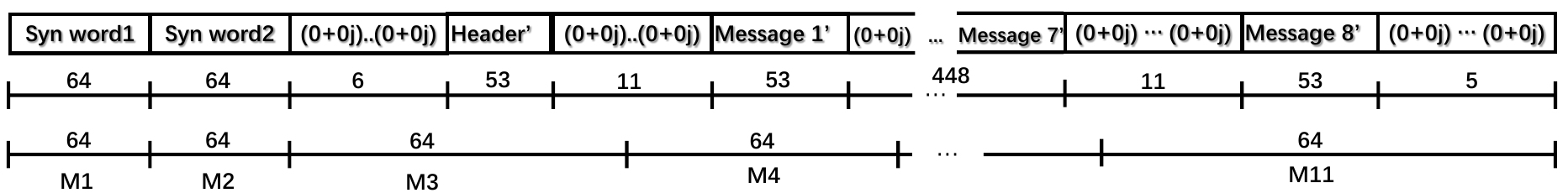}
  \caption{Complex number stream after the OFDM carrier manipulator block}\label{fig:Carrier_Manipulator}
\end{figure*}

The OFDM-carrier-allocator block maps the stream of $432$ complex numbers into $11$ complex vectors, which are shown in Figure \ref{fig:Carrier_Manipulator}. The complex vectors are labeled as $M_i$ for $i=1,2,\ldots,11$, and each vector contains 64 complex numbers as 64 subcarriers, where $M_1$ and $M_2$ are two synchronization words. Additionally, each header prime and each message prime in Figure \ref{fig:Carrier_Manipulator} come from the header and message data after inserted $4$ pilot carriers and $0$ DC subcarrier. Namely, the pilot complex numbers $[1, 1, 1, -1]$ are inserted into the subcarriers $[-21, -7, 7, 21]$ respectively for each of $64$ subcarriers. Furthermore, the subcarriers from $-32$ to $-27$ and from $27$ to $31$ and subcarrier $0$ are set to be complex value zeros. Thus, the format of subcarriers exactly matches the IEEE 802.11 standard in Figure \ref{fig:OFDM_802.11a_Keysight}.

The size of the IFFT block is set to 64 so that it can manipulate 64 complex subcarriers at a time. More specifically, it converts 64 complex numbers that are discrete samples in the frequency domain to 64 complex numbers that are discrete samples in the time domain.

The cyclic prefix  block inserts a cyclic prefix (CP) consisting of $16$ complex numbers  at the beginning of every stream of $64$ complex numbers. The prefix is the copy of the last 16 complex numbers out of the $64$ complex numbers.

The multiply-const block is used to multiply each input complex number by a constant number in order to adjust the gain of signals. The constant number is set from $0.01$ to $0.03$ for the implementation by using USRPs. In other words, if the constant number is lower than 0.01, then the signal-to-noise ratio (SNR) will be too small. In contrast, if the constant number is higher than $0.03$, the USRP will be saturated for a high SNR. 
The SDR will receive a high bit-error-rate (BER) for both situations. 
Note that this constant number is the reference number for the USRP devices which may not be linearly proportion to the sending signal's power, and its range is not accurate for each USRP device. 
Finally, the tag-gate block is used to remove the tag which is an internal variable passed by blocks.

The USRP-sink block in the GNU radio companion provides an interface to setup the parameters of the USRP device, which has parameters, namely IP address, center frequency and sample rate. In our experiment, the USRP-sink block is used to set the parameters for the USRP sender whose IP address is set to be $addr=\texttt{192.168.10.2}$ and the center frequency is set to 892 MHz.

\begin{figure*}[ht]
  \centering
   \vspace{-0.3cm}
  \includegraphics[width=\linewidth]{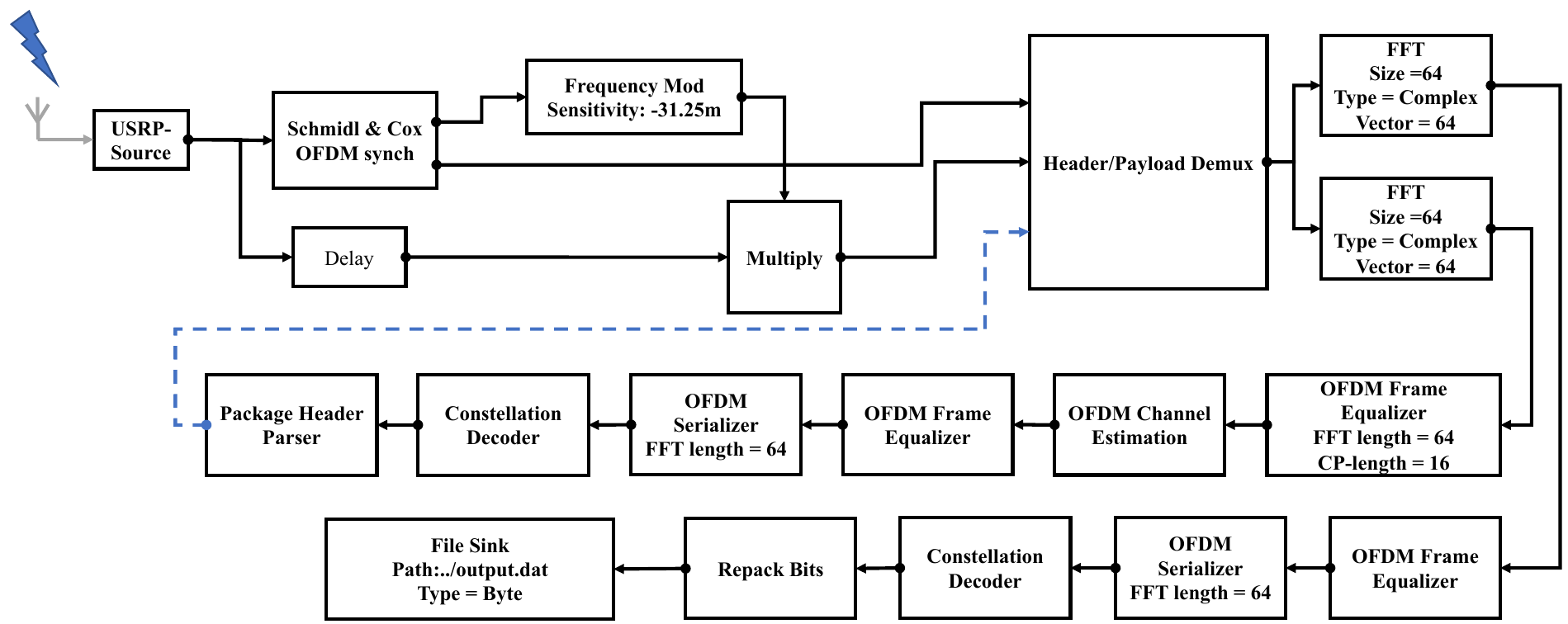}
  \caption{A block diagram of the OFDM receiver}\label{fig:OFDM_Receiver}
  \vspace{-0.5cm}
\end{figure*}

\noindent\textbf{IEEE 802.11a OFDM Receiver.} The OFDM receiver is shown in Figure \ref{fig:OFDM_Receiver}. The OFDM demodulation procedures include a header detector, FFT, frame equalizer, OFDM serializer, underlying demodulation and repack are put in one block which used Schmidl OFDM Synchronization block given by \cite{Schmidl97} in order to increase the efficiency of frequency and timing synchronization. The file-sink block in Figure \ref{fig:OFDM_Receiver} is used to save the demodulated message bytes and to get BER by comparing with the original sending data. 
The USRP-source block for the receiver  has the same parameter as of the sender, except the IP, which is $addr= \texttt{192.168.60.2}$.

\subsection{Experiment Setup for the IEEE 802.11X and CoAP Handshaking Mutual Authentication}\label{subsec-Impl}
We implement the 4-way and 6-way handshake protocols in Figures~\ref{4-way} and~\ref{6-way}  using Atmel Studio, IAR, and two USRP N210 devices to generate, send and receive messages.
We record the time for transmitting each 96 bytes message in the tag debug block and calculate the time of generating tags from IAR. The OFDM system is a tagged system, the sending message has to be the multiple of $96$ bytes or the last $96$ bytes won't be sent. Therefore, we pad zeros after $ANonce$, $SNonce||MIC_A$, $ANonce\|MIC_S$, $MIC_{all}$, $HelloClient$, $HelloServer$, etc. to reach $96$ bytes. Therefore, the 4-way or 6-way handshake protocols transmit $4 \times 96 =384$ or $6 \times 96 = 576$ bytes.

We implement the {\kdf} and {\mic} algorithms described in Section~\ref{sec-Implementation} for {\ace}, {\spix} and {\wage} and record the key derivation and MIC computation times in the IAR and Atmel Studio. 
The 4-way and 6-way transmission timings are captured in the SDR. Finally, the total time for the 4-way or 6-way handshake protocol is the sum of the 4-way or 6-way transmission time, the session key generation time and the MIC computation time. 

To assess the efficiency of the data protection phase, we consider  associated data and plaintext messages of two different lengths. 
In the first case, we choose no associated data and a 1024-bit plaintext message. For the second case, we choose an AD is of 128 bits and a plaintext message of $1024$ bits.
We record the execution times for {\ace}, {\spix} and {\wage} in the AEAD mode in the USRP interface.

\section{Experiment Results and Comparisons}\label{sec-Results}
%
\noindent{\bf Performance of the IEEE802.11X and CoAP protocols.} 
Table \ref{Tab:Performance_KDF_MIC} presents the execution time of the IEEE802.11X and CoAP handshake protocols where the execution times for the communication module and cryptographic functionalities are separately shown for a clear understanding. 
In our experiments, the frame size of the USRP is 1472 bytes, and the average frame rate of USRP is about 16.82 Kbps.
In Table~\ref{Tab:Performance_KDF_MIC}, the 4-way-Tx-time and 6-way-Tx-time is the 4-way and 6-way transmission time for the handshake protocols, which is the time required by the USRP. 
It takes about 700 and 1050 milliseconds (ms) to transmit the  messages in the 4-way and 6-way communications, respectively.  
The ``Gen-time'' is the execution time for one {\kdf} or {\mic} according to Constructions~\ref{const:KDF} and~\ref{const:MIC}. 
The ``WiFi-Auth'' (``CoAP-Auth'') in Table \ref{Tab:Performance_KDF_MIC} is the total time to complete a mutual authentication, which includes the 4-way (or 6-way) transmission time and the execution times of 2 {\kdf}s and 6 {\mic}s (or 2 {\kdf}s, 2 hash, 6 {\mic}s) computations. 
For instance, with {\ace}, {\spix}, and {\wage}, the execution time to complete the handshake protocol is about 2831, 2966, and 2808 ms, respectively. 

\begin{table*}[h]
\centering
\caption{Performance of {\kdf} and {\mic} on microcontrollers at a clock frequency of 16 MHz and time for the IEEE 802.11X  and CoAP handshake mutual authentication and key establishment protocols}
\scalebox{0.55}{
\begin{tabular}{||c|c|c|c|c|r|r|r|c|c|c|c||}
\hline
\multirow{2}{*}{\begin{tabular}[c]{@{}c@{}}{\bf Cryptographic}\\ {\bf Primitives}\end{tabular}} & \multirow{2}{*}{{\bf Platform}} & \multirow{2}{*}{{\bf Function}} & \multicolumn{2}{c|}{\begin{tabular}[c]{@{}c@{}}{\bf Memory usage}\\ {[}Bytes{]}\end{tabular}} & \multirow{2}{*}{\begin{tabular}[c]{@{}c@{}}{\bf Setup}\\ {[}Cycles{]}\end{tabular}} & \multirow{2}{*}{\begin{tabular}[c]{@{}c@{}}{\bf Throughput}\\ {[}Kbps{]}\end{tabular}} & \multirow{2}{*}{\begin{tabular}[c]{@{}c@{}}{\bf Gen-time}\\ {[}ms{]}\end{tabular}} & \multirow{2}{*}{\begin{tabular}[c]{@{}c@{}}{\bf 4-way-Tx-time}\\ {[}ms{]}\end{tabular}} & \multirow{2}{*}{\begin{tabular}[c]{@{}c@{}}{\bf WiFi-Auth}\\ {[}ms{]}\end{tabular}} & \multirow{2}{*}{\begin{tabular}[c]{@{}c@{}} {\bf 6-way-Tx-time}\\ {[}ms{]}\end{tabular}}&\multirow{2}{*}{\begin{tabular}[c]{@{}c@{}}{\bf CoAP-Auth$^{\dagger}$}\\ {[}ms{]}\end{tabular}}\\ \cline{4-5}
 &  &  & SRAM & Flash &  &  &  &  & &&\\\hline\hline
\multirow{6}{*}{\spix} & \multirow{2}{*}{\begin{tabular}[c]{@{}c@{}}8-bits\\ ATmega128\end{tabular}} & \kdf & 175 & 1,586 & 705,314 & 23.23 & 44.08 & \multirow{2}{*}{700} & \multirow{2}{*}{3,176.64} &\multirow{2}{*}{1060}&\multirow{2}{*}{6,857}\\ \cline{3-8}
 &  & \mic & 175 & 1,590 & 833,251 & 2.46 & 52.08 &  &  & &\\ \cline{2-12}
 & \multirow{2}{*}{\begin{tabular}[c]{@{}c@{}}16-bits\\ MSP430F2013\end{tabular}} & \kdf & 50 & 1,562 & 286,679 & 57.15 & 17.92 & \multirow{2}{*}{690} & \multirow{2}{*}{2,912.84} &\multirow{2}{*}{1050}&\multirow{2}{*}{6,502}\\ \cline{3-8}
 &  & \mic & 50 & 1,580 & 338,106 & 6.06 & 21.13 &  &  &&\\ \cline{2-12}
 & \multirow{2}{*}{\begin{tabular}[c]{@{}c@{}}32-bits\\ LM3S9D96\end{tabular}} & \kdf & 408 & 1,230 & 59,140 & 277.04 & 3.70 & \multirow{2}{*}{700} & \multirow{2}{*}{2,831.58} &\multirow{2}{*}{1050}&\multirow{2}{*}{6,342}\\ \cline{3-8}
 &  & \mic & 408 & 1,294 & 69,770 & 29.35 & 4.36 &  &  &&\\ \hline
\multirow{4}{*}{\ace} & \multirow{2}{*}{\begin{tabular}[c]{@{}c@{}}16-bits\\ MSP430F2013\end{tabular}} & \kdf & 330 & 1,720 & 550,752 & 29.75 & 34.42 & \multirow{2}{*}{710} & \multirow{2}{*}{3,089.68} &\multirow{2}{*}{1040}&\multirow{2}{*}{6,567}\\ \cline{3-8}
 &  & \mic & 330 & 1,740 & 551,016 & 3.72 & 34.44 &  &  &&\\ \cline{2-12}
 & \multirow{2}{*}{\begin{tabular}[c]{@{}c@{}}32-bits\\ LM3S9D96\end{tabular}} & \kdf & 599 & 1,826 & 102,762 & 159.44 & 6.42 & \multirow{2}{*}{730} & \multirow{2}{*}{2,966.56} &\multirow{2}{*}{1070}&\multirow{2}{*}{6,481}\\ \cline{3-8}
 &  & \mic & 599 & 1,790 & 102,746 & 19.93 & 6.42 &  &  &&\\ \hline
\multirow{6}{*}{\wage} & \multirow{2}{*}{\begin{tabular}[c]{@{}c@{}}8-bits\\ ATmega128\end{tabular}} & \kdf & 808 & 4,448 & 139,478 & 117.47 & 8.72 & \multirow{2}{*}{710} & \multirow{2}{*}{2,903.22} &\multirow{2}{*}{1060}&\multirow{2}{*}{6,443}\\ \cline{3-8}
 &  & \mic & 808 & 4,476 & 139,309 & 14.70 & 8.71 &  &  &&\\ \cline{2-12}
 & \multirow{2}{*}{\begin{tabular}[c]{@{}c@{}}16-bits\\ MSP430F2013\end{tabular}} & \kdf & 46 & 4,518 & 166,993 & 98.11 & 10.44 & \multirow{2}{*}{720} & \multirow{2}{*}{2,955.78} &\multirow{2}{*}{1050}&\multirow{2}{*}{6,399}\\ \cline{3-8}
 &  & \mic & 46 & 4,538 & 166,865 & 12.27 & 10.43 &  &  &&\\ \cline{2-12}
 & \multirow{2}{*}{\begin{tabular}[c]{@{}c@{}}32-bits\\ LM3S9D96\end{tabular}} & \kdf & 3,084 & 6,278 & 107,071 & 153.02 & 6.69 & \multirow{2}{*}{690} & \multirow{2}{*}{2,808.54} &\multirow{2}{*}{1060}&\multirow{2}{*}{6,424}\\ \cline{3-8}
 &  & \mic & 3,084 & 6,326 & 106,977 & 19.14 & 6.69 &  &  &&\\ \hline
\end{tabular}\label{Tab:Performance_KDF_MIC}
}
\vspace{0.3cm}
 {$\dagger$ For CoAP-Auth, {\spix} and {\wage} provide the 112-bit security due to the security of {\spix} and {\wage} hash.}
\end{table*}


\noindent{\bf Performance of the data protection protocol.}
The execution time in the data protection protocol includes the execution time of encrypting a plaintext message and producing a tag using an AEAD algorithm and the transmission time of sending the ciphertext and the tag. 
The transmission time is recorded in the SDR for transmitting 1024-bit ciphertext and 128 bit tag from one USRP to another USRP, which takes about 1060 ms to transmit the ciphertext and tag.  
Tables~\ref{Tab:SPIX}, \ref{Tab:ACE}, and \ref{Tab:WAGE} present the transmission time and the AEAD execution times of {\spix}, {\ace}, and {\wage}, respectively. 
As in all three AEAD algorithms, the permutation is at the core, we present its performance, the memory usage, setup, and throughput results. 
%
Note that ``Gen-time'' in Tables~\ref{Tab:Performance_KDF_MIC}, \ref{Tab:SPIX}, \ref{Tab:ACE} and \ref{Tab:WAGE} is the time to one execution of the respective AEAD algorithm. 
For example, the total time to complete the data protection protocol with a plaintext message of $1024$ bits and an AD of 128 bits is about 1304 ms using {\spix} on ATmega128. 
\begin{table}[h]
\centering
\caption{Performance of SPIX AE mode on microcontrollers at a clock frequency of 16 MHz IEEE 802.11i data protection protocol}
\scalebox{0.7}{
\begin{tabular}{||c|c|c|c|c|c|c|c|c||}
\hline
\multirow{2}{*}{Cryptographic}                                  & \multirow{2}{*}{Platform}                                                           & \multicolumn{2}{c|}{\begin{tabular}[c]{@{}c@{}}Memory usage \\ {[}Bytes{]}\end{tabular}} & \multirow{2}{*}{\begin{tabular}[c]{@{}c@{}}Setup\\ {[}Cycles{]}\end{tabular}} & \multirow{2}{*}{\begin{tabular}[c]{@{}c@{}}Throughput\\ {[}Kbps{]}\end{tabular}} & \multirow{2}{*}{\begin{tabular}[c]{@{}c@{}}Gen-time\\ {[}ms{]}\end{tabular}}& \multirow{2}{*}{\begin{tabular}[c]{@{}c@{}}Tx-time \\ {[}ms{]}\end{tabular}} \\ \cline{3-4}
                                                                &                                                                                  & SRAM                                       & Flash                                       &                                                                               &                                                                                  &
                                                                &                                                                                \\ \hline\hline
\multirow{3}{*}{SPIX Perm-18}  & \begin{tabular}[c]{@{}c@{}}8-bits\\ ATmega128\end{tabular}  & 161 & 1262 & 128377 & 31.91   & 8.02 & \multirow{3}{*}{N/A}\\ \cline{2-7}
& \begin{tabular}[c]{@{}c@{}}16-bits\\ MSP430F2013\end{tabular}  & 24 & 1409 & 52294 & 78.33  & 3.27 &\\ \cline{2-7}
& \begin{tabular}[c]{@{}c@{}}32-bits\\ LM3S9D96\end{tabular}  & 352 & 946 & 10900 & 375.78 & 0.68 &\\ \hline

\multirow{3}{*}{\begin{tabular}[c]{@{}c@{}}SPIX-AE\\ ($l_{AD}=0$,  \\ $l_M=16$) \end{tabular}}  & \begin{tabular}[c]{@{}c@{}}8-bits\\ ATmega128\end{tabular}  & 175 & 1550 & 1667042 & 9.83 & 104.19 & 1,060\\ \cline{2-8}
& \begin{tabular}[c]{@{}c@{}}16-bits\\MSP430F2013\end{tabular}  & 50 & 1845 & 677818  & 24.17 & 42.36 & 1,080\\ \cline{2-8}

& \begin{tabular}[c]{@{}c@{}}32-bits\\LM3S9D96\end{tabular}  & 408 & 1210 & 139569 & 117.39 & 8.72 & 1,050\\ \hline
\multirow{3}{*}{\begin{tabular}[c]{@{}c@{}}SPIX-AE\\ ($l_{AD}=2$,\\ $l_M=16$)\end{tabular}}  & \begin{tabular}[c]{@{}c@{}}8-bits\\ATmega128\end{tabular} & 175 & 1644 & 1795322 & 9.13 & 112.21 & 1,080\\ \cline{2-8}
& \begin{tabular}[c]{@{}c@{}}16-bits\\MSP430F2013\end{tabular} & 50 & 1891 & 730340 & 22.43 & 45.65 & 1,050\\ \cline{2-8}
& \begin{tabular}[c]{@{}c@{}}32-bits\\LM3S9D96\end{tabular} & 424 & 1326 & 150313 & 109.00 & 9.39 &1,070\\ \hline
\end{tabular}\label{Tab:SPIX}
}
\end{table}
\begin{table}[h]
\centering
\caption{Performance of ACE AE and Hash modes on microcontrollers at a clock frequency of 16 MHz IEEE 802.11i data protection protocol}
\scalebox{0.7}{
\begin{tabular}{||c|c|c|c|c|c|c|c||}
\hline
\multirow{2}{*}{Cryptographic}                                  & \multirow{2}{*}{Platform}                                                           & \multicolumn{2}{c|}{\begin{tabular}[c]{@{}c@{}}Memory usage \\ {[}Bytes{]}\end{tabular}} & \multirow{2}{*}{\begin{tabular}[c]{@{}c@{}}Setup\\ {[}Cycles{]}\end{tabular}} & \multirow{2}{*}{\begin{tabular}[c]{@{}c@{}}Throughput\\ {[}Kbps{]}\end{tabular}} & \multirow{2}{*}{\begin{tabular}[c]{@{}c@{}}Gen-time\\ {[}ms{]}\end{tabular}} & \multirow{2}{*}{\begin{tabular}[c]{@{}c@{}}Tx-time\\ {[}ms{]}\end{tabular}} \\ \cline{3-4}
                                                                &                                                                                  & SRAM                                       & Flash                                       &                                                                               &                                                                                  &
                                                                &                                                                               \\ \hline\hline
\multirow{2}{*}{ACE Perm}  & \begin{tabular}[c]{@{}c@{}}16-bits\\ MSP430F2013\end{tabular}  & 304 & 1456 & 69440 & 73.73 & 4.34& \multirow{2}{*}{N/A}\\ \cline{2-7}
 & \begin{tabular}[c]{@{}c@{}}32-bits\\ LM3S9D96\end{tabular}  & 523 & 1598 & 13003 & 393.76 & 0.81& \\ \hline

\multirow{2}{*}{\begin{tabular}[c]{@{}c@{}}ACE-AE\\ ($l_{AD}=0$,\\ $l_M=16$)\end{tabular}}  & \begin{tabular}[c]{@{}c@{}}16-bits\\MSP430F2013\end{tabular}  & 330 & 1740 & 1445059 & 11.34 & 90.32 &1,060\\ \cline{2-8}
& \begin{tabular}[c]{@{}c@{}}32-bits\\LM3S9D96\end{tabular}  & 559 & 1790 & 269341 & 60.83 & 16.83 &1,070\\ \hline

\multirow{2}{*}{\begin{tabular}[c]{@{}c@{}}ACE-AE\\ ($l_{AD}=2$,\\ $l_M=16$)\end{tabular}}  & \begin{tabular}[c]{@{}c@{}}16-bits\\MSP430F2013\end{tabular}  & 330 & 1786 & 1582892 & 10.35 & 98.93 &1,080\\ \cline{2-8}
& \begin{tabular}[c]{@{}c@{}}32-bits\\LM3S9D96\end{tabular}  & 559 & 1858 & 294988 & 55.54 & 18.44 &1,080\\ \hline

\multirow{2}{*}{\begin{tabular}[c]{@{}c@{}}ACE-Hash\\ ($l_M=2$, \\ $j$ =4)\end{tabular}}  & \begin{tabular}[c]{@{}c@{}}16-bits\\MSP430F2013\end{tabular}  & 330 & 1682 & 413056 & 4.96 & 25.82 &\multirow{3}{*}{N/A}\\ \cline{2-7}
&  \begin{tabular}[c]{@{}c@{}}32-bits\\LM3S9D96\end{tabular}  & 559 & 1822 & 77114 & 26.56 & 4.82 &\\ \cline{1-7}

\multirow{2}{*}{\begin{tabular}[c]{@{}c@{}}ACE-Hash\\ ($l_M=16$, \\ $j$ =4)\end{tabular}}  & \begin{tabular}[c]{@{}c@{}}16-bits\\MSP430F2013\end{tabular}  & 330 & 1684 & 1375672 & 11.91 & 85.98 &\\ \cline{2-7}
& \begin{tabular}[c]{@{}c@{}}32-bits\\LM3S9D96\end{tabular}  & 559 & 1822 & 256524 & 63.87 & 16.03 &\\ \hline

\end{tabular}\label{Tab:ACE}
}
\vspace{-0.3cm}
\end{table}

\noindent{\bf Comparing with AES.} 
We compare the throughput of the {\spix}, {\ace} and {\wage} permutations with the AES-128 permutation.  
The implementation results of AES on 8-bit AVR microcontrollers (written in C) from \cite{Meiser08} shows that the throughput of the AES-128 permutation is $\frac{10180\times8\times2}{1000} =162.880$ Kbps when the CPU frequency is set to 16 MHz. 
When we compare the results of AES-128 with {\wage}, our implementation results of {\wage} gives a higher throughput, which is $217.98$ Kbps, on the same 8-bit microcontroller platforms. 
Moreover, {\spix} and {\ace} permutations give higher throughput on 32-bit microcontrollers which are $393.76$ Kbps and $286.78$ Kbps, respectively. 

When AES is written in the assembly language, the throughput of the AES-128 permutation is $\frac{43671\times8\times 2}{1000} = 698.74$ Kbps, which is higher than that of our implementations. 
However, the internal state size of {\spix}, {\ace} and {\wage} are 256, 320 and 259 bits, respectively, which are more than twice as much as that of AES-128. 
\begin{table}[h]
\centering
\caption{Performance of WAGE AE mode  on microcontrollers at a clock frequency of 16 MHz in IEEE 802.11i data protection protocol}
\scalebox{0.7}{
\begin{tabular}{||c|c|c|c|c|c|c|c||}
\hline
\multirow{2}{*}{Cryptographic}                                  & \multirow{2}{*}{Platform}                                                           & \multicolumn{2}{c|}{\begin{tabular}[c]{@{}c@{}}Memory usage \\ {[}Bytes{]}\end{tabular}} & \multirow{2}{*}{\begin{tabular}[c]{@{}c@{}}Setup\\ {[}Cycles{]}\end{tabular}} & \multirow{2}{*}{\begin{tabular}[c]{@{}c@{}}Throughput\\ {[}Kbps{]}\end{tabular}} & \multirow{2}{*}{\begin{tabular}[c]{@{}c@{}}Gen-time\\ {[}ms{]}\end{tabular}} & \multirow{2}{*}{\begin{tabular}[c]{@{}c@{}}Tx-time\\ {[}ms{]}\end{tabular}} \\ \cline{3-4}
                                                                &                                                                                  & SRAM                                       & Flash                                       &                                                                               &                                                                                  &
                                                                &                                                                                \\ \hline\hline
\multirow{3}{*}{WAGE Perm}                      & \begin{tabular}[c]{@{}c@{}}8-bits\\ ATmega128\end{tabular}    & 802         & 4132                  & 19011                 & 217.98       & 1,190                             &   \multirow{3}{*}{N/A}\\ \cline{2-7}
& \begin{tabular}[c]{@{}c@{}}16-bits\\ MSP430F2370\end{tabular} & 4 & 5031 & 23524 & 176.16 & 1.47&                       \\ \cline{2-7}
& \begin{tabular}[c]{@{}c@{}}32-bits\\ LM3S9D96\end{tabular} & 3076 & 5902 & 14450 & 286.78 & 0.9 &                      \\ \hline
\multirow{3}{*}{\begin{tabular}[c]{@{}c@{}}WAGE-AE\\ ($l_{AD}=0$, \\$l_M=16$)\end{tabular}}  & \begin{tabular}[c]{@{}c@{}}8-bits\\ ATmega128\end{tabular} & 808 & 4416 & 362888 & 45.15 & 22.68 &1,080\\ \cline{2-8}
& \begin{tabular}[c]{@{}c@{}}16-bits\\ MSP430F2370\end{tabular} & 46 & 5289 & 433105 & 37.83 & 27.07 &1,090\\ \cline{2-8}
& \begin{tabular}[c]{@{}c@{}}32-bits\\ LM3S9D96\end{tabular} & 3084 & 6230 & 278848 & 58.76 & 17.43 &1,060\\ \hline
\multirow{3}{*}{\begin{tabular}[c]{@{}c@{}}WAGE-AE\\ ($l_{AD}=2$, \\$l_M=16$)\end{tabular}}  & \begin{tabular}[c]{@{}c@{}}8-bits\\ ATmega128\end{tabular} & 808 & 4502 & 397260 & 41.24 & 24.83 &1,050\\ \cline{2-8}
 & \begin{tabular}[c]{@{}c@{}}16-bits\\ MSP430F2370\end{tabular} & 46 & 5339 & 474067 & 34.56 & 29.63 &1,060\\ \cline{2-8}
 & \begin{tabular}[c]{@{}c@{}}32-bits\\ LM3S9D96\end{tabular}   &  3084 &  6354 & 305284 & 53.67      & 19.08 &1,060\\ \hline
\end{tabular}\label{Tab:WAGE}
}
\vspace{-0.25cm}
\end{table}

\noindent{\bf Scaling up the Speed for WiFi System.} 
Recall that, in our experiment, the USRP transmission rate is about 16.82 Kbps.  
However, the real WiFi systems have a transmission rate in the range of 50 Mbps and 320 Mbps at distance of 100m from devices to an access point, which is much higher than that of the USRP.

We compute the equivalent 4-way transmission time for the WiFi system by scaling the 4-way transmission time of the USRP. 
For the 4-way transmission time with {\spix} that takes about 700 ms, the equivalent transmission time for the WiFi system at a transmission rate of 50Mbps is $\frac{0.7 \times 16.82}{50000} = 0.235 $ ms. 
Similarly it can be computed for the 6-way transmission time.  
Therefore, from Table~\ref{Tab:Performance_KDF_MIC} we can observe that the execution time for the cryptographic operations is the dominating factor in the 4-way or 6-way handshake protocol. 
The data transmission in the 5G or satellite communications \cite{jiang19satellite} is much more expensive than the WiFi transmission.

\section{Conclusion and Future Work}\label{sec-concl}
In this paper, we implemented the IEEE 802.11X 4-way and CoAP 6-way handshake protocols and the IEEE 802.11a physical layer OFDM transmission protocol in software defined radio and embed the handshake protocols into the IEEE 802.11a protocol to simulate the 4-way and 6-way handshake modulation and communication. 
We have proposed the construction of {\kdf} and {\mic} algorithms and implemented three LWC schemes, namely {\ace}, {\spix} and {\wage} including  {\kdf} and {\mic} on three different types of microcontrollers. 
The experimental results for two IoT authentication protocols namely IEEE 802.11X and CoAP and the data protection protocols are reported. 
Our results show that for authenticated encryption all three ciphers achieved the highest throughput on Cortex-M3. 
A comparison of our implementations of {\ace}, {\spix} and {\wage} with AES-128 on the 8-bit ATmega128 platform is done. 
 
In the current IEEE 802.11i (as well as amended one) protocol, the cipher suite has only AES and the data protection protocol has the CCMP and GCMP schemes. Our experimental results for the cryptographic functionalities and the radio communication phase in SDR provide the insight for the design choices for IoT devices connected through the WiFi to Internet. Our experimental setup will facilitate further experimental research on anti jamming, location service attacks, and entry point intrusion attacks.

In cellular systems, 5G  will adopt 4G-LTE's authentication and key agreement (AKA) protocol. After a successful  full authentication, it may execute multiple local authentications. 
In the local authentication, the  AKA protocol running between a wireless device and the mobility management entity is a 2-round sequence number-based authentication protocol. 
As a future work, we extend our experiment setup for the SDR  and the LWC schemes to provide performance evaluations of upcoming 5G security mechanisms and other new IoT protocols. 

\bibliographystyle{plain}
\bibliography{LWC}

\vspace{0.2cm}
\begin{center}
 \appendix{\Large \bf Appendix}
\end{center}

\section{Orthogonal frequency-division multiplexing (OFDM) system}\label{sec:OFDM}
In this subsection, we introduce  the basic structure of the orthogonal frequency-division multiplexing (OFDM) system. In detail, it gives basic concepts of underlying modulation, inverse discrete fourier transform (IDFT), discrete fourier transform (DFT), the orthogonality and cyclic prefix (CP) for a general OFDM system.

In \cite{Chang66}, the authors demonstrated a communication scheme called OFDM to transmit multiple messages simultaneously on a linear bandlimited channel without involving intersymbol interference (ISI) and inter-channel interference. The total bandwidth $W$ has been divided into multiple sub-channels, and those sub-channels are overlapping with each other one by one. However, they will not affect each other during the transmission in a linear bandlimited channel due to the orthogonality of the subcarriers. The basic model of OFDM system contains serial-to-parallel conversion, underlying modulation, N-inverse fast fourier transform (N-IFFT) and digital-to-analog convertor (DAC) conversion at the sender side. At the receiver side, it contains analog-to-digital convertor (ADC) conversion, N-fast fourier transform (N-FFT), parallel-to-serial conversion and underlying demodulation (see \cite{Andreabook}).  From \cite{Vlaovi16} and \cite{Steendam13}, pilot symbols and pilot carriers are used for channel estimation. The number of subcarriers $N$ is equal to the size of the IFFT and FFT, and each subcarrier is orthogonal to each other.

The underlying modulation is also called sub-carrier modulation. The process of underlying modulation is done before the serial-to-parallel conversion. The purpose of the underlying modulation is to map the input bits into constellation in complex domain. The underlying modulations include binary phase-shift keying (BPSK), quadrature phase-shift keying (QPSK), quadrature amplitude modulation (QAM) and so on. The selection of those modulations depends on the channel condition and the communication regulation. The bit error probability of M-ary phase-shift keying (MPSK) is given in Equation (\ref{Equ:BER_MPSK}) \cite{Andreabook}. 
\begin{equation}\label{Equ:BER_MPSK}
P_b \approx \frac{2}{\log_2{M}} Q(\sqrt{\frac{2E_b}{N_0} \log_2 M sin(\frac{\pi}{M}})).
\end{equation} 
Under the same signal to noise (SNR), $\frac{E_b}{N_0}$, increasing the value $M$ will increase the bit error probability.

IFFT is a significant part in the OFDM system. The IDFT is shown in Equation (\ref{Equ:IFFT}).
\begin{equation}\label{Equ:IFFT}
s_i=\frac{1}{\sqrt{N}} \sum_{k=0}^{N-1} S_k e^{\frac{j2\pi ik}{N}} , i, k=0,1,\cdots,N-1.
\end{equation}
The IDFT is used at the OFDM sender to convert frequency domain samples to time domain samples. The IFFT has a lower complexity to get the time domain samples for the realization purpose in hardware.

The DFT is shown in Equation (3).
\begin{equation}\label{Equ:FFT}
S_i=\frac{1}{\sqrt{N}} \sum_{k=0}^{N-1} s_k e^{-\frac{j2\pi ik}{N}} , i, k=0,1,\cdots,N-1.
\end{equation}
The IDFT is used at the OFDM receiver to calculate the frequency domain samples from the time domain samples. The FFT is the low complexity method to calculate the DFT in hardware.

Orthogonality is a word to demonstrate that the frequency domain signals do not affect each other and the product integral between their time domain signals is zero. In fact, the samples before the IFFT in OFDM system are viewed as discrete frequency samples, and the IFFT will convert them to discrete time samples. Cyclic Prefix is used to reduce the ISI \cite{Andreabook}. According to \cite{Ohno07}, if the duration of CP is longer than channel delay spread, the ISI will be completely removed. The reason that uses CP instead of using padding zeros is to avoid involving DC offset which increases the BER a lot \cite{Huang06}. The prefix interval generally will be $N/4$ which is 16 when $N$ is equal to 64 in our case.

\end{document}